\documentclass[preprint2]{aastex} %nice

\usepackage{amsmath}
\usepackage[]{natbib}
\usepackage{multirow}
\usepackage{soul} %provides \st

\newcommand{\mdot}{\ensuremath{\dot{M}}}
\newcommand{\uJy}{$\mu$Jy~}

\newcommand{\ditto}{$\prime\prime$}
\newcommand{\ergshz}{\ensuremath{\mathrm{erg}\,\mathrm{s}^{-1}\mathrm{Hz}^{-1}}}
\newcommand{\moyr}{\ensuremath{\mathrm{M}_\sun\mathrm{yr}^{-1}}}
\newcommand{\msun}{\ensuremath{\mathrm{M}_\sun}}
\newcommand{\kms}{\ensuremath{\mathrm{km\,s}^{-1}}}
\newcommand{\e}[1]{\times 10^{#1}}

\slugcomment{.} %the removal of this text causes major formatting issues.

\shorttitle{Visibility stacking and SNIa radio emission.}
\shortauthors{Hancock et al.}

\begin{document}

%% LaTeX will automatically break titles if they run longer than
%% one line. However, you may use \\ to force a line break if
%% you desire.

\title{Visibility stacking in the quest for SNIa radio emission.}

%% Use \author, \affil, and the \and command to format
%% author and affiliation information.
%% Note that \email has replaced the old \authoremail command
%% from AASTeX v4.0. You can use \email to mark an email address
%% anywhere in the paper, not just in the front matter.
%% As in the title, use \\ to force line breaks.

\author{P. Hancock\altaffilmark{1,2}, B. M. Gaensler\altaffilmark{1,2} and T. Murphy\altaffilmark{1,2,3}}
\affil{$^1$Sydney Institute for Astronomy (SIfA), School of Physics, The University of Sydney, NSW 2006, Australia}
\affil{$^2$ARC Centre of Excellence for All--sky Astrophysics (CAASTRO)}
\affil{$^3$School of Information Technologies, The University of Sydney, NSW 2006, Australia}
\email{Paul.Hancock@sydney.edu.au}

%% Notice that each of these authors has alternate affiliations, which
%% are identified by the \altaffilmark after each name.  Specify alternate
%% affiliation information with \altaffiltext, with one command per each
%% affiliation.

%\altaffiltext{1}{bla bla bla}

%% Mark off your abstract in the ``abstract'' environment. In the manuscript
%% style, abstract will output a Received/Accepted line after the
%% title and affiliation information. No date will appear since the author
%% does not have this information. The dates will be filled in by the
%% editorial office after submission.

\begin{abstract}
We describe the process of stacking radio interferometry visibilities to form a deep composite image and its application to the observation of transient phenomena. We apply ``visibility stacking'' to 46 archival Very Large Array observations of nearby type Ia supernovae (SNeIa). This new approach provides an upper limit on the SNIa ensemble peak radio luminosity of $1.2\e{25}\ergshz$ at 5~GHz, which is 5--10 times lower than previously measured. This luminosity implies an upper limit on the average companion stellar wind mass loss rate of $1.3\e{-7}\moyr$. This mass loss rate is consistent with the double degenerate scenario for SNeIa and rules out intermediate and high mass companions in the single degenerate scenario. In the era of time domain astronomy, techniques such as visibility stacking will be important in extracting the maximum amount of information from observations of populations of short lived events.
\end{abstract}

%% Keywords should appear after the \end{abstract} command. The uncommented
%% example has been keyed in ApJ style. See the instructions to authors
%% for the journal to which you are submitting your paper to determine
%% what keyword punctuation is appropriate.


%% From the front matter, we move on to the body of the paper.
%% In the first two sections, notice the use of the natbib \citep
%% and \citet commands to identify citations.  The citations are
%% tied to the reference list via symbolic KEYs. The KEY corresponds
%% to the KEY in the \bibitem in the reference list below. We have
%% chosen the first three characters of the first author's name plus
%% the last two numeral of the year of publication as our KEY for
%% each reference.

\section{Introduction}
Type Ia supernovae (SNeIa) have been important to astronomy since their identification as standard candles, and subsequent use in measureing the accelerating expansion of the universe \citep{Riess1998,Perlmutter1999}. The current uncertainties in measuring cosmological parameters are dominated by systematic rather than statistical errors \citep{Neill2006}. Hence, in order to better confine the cosmological parameters it is important to understand the origin and evolution of SNeIa \citep{Hillebrandt2000}.

Recent work has been of an empirical nature, characterizing the differences between SNIa light curves as a function of environment \citep{Hamuy2000}, redshift, and metallicity \citep{Timmes2003}. Another approach is to understand the physics of a SNIa explosion from a fundamental level so that the shape and luminosity of the light curve, and its subsequent evolution, can be predicted and understood more completely. One of the key parameters affecting the light curve evolution is the amount of circumstellar material (CSM) that is present.

As well as causing optical extinction, the expected CSM provides a medium with which the supernova (SN) ejecta can interact and produce radio synchrotron emission. This process observed in type~II and Ib/c SN. The spectral and temporal distribution of this emission has been used to explore the environment and history of the progenitor systems \citep{Montes1998}. Despite the usefulness of this approach in characterizing the progenitor system, the radio detection of SNeIa has remained elusive.

The mechanism by which SNIa white dwarf progenitors obtain a critical mass of $\sim1.4\msun$ is thought to be either accretion from, or merger with, a binary companion. In the single degenerate (SD) scenario, a CO white dwarf accretes material from a companion star which is losing mass via Roche--lobe overflow or strong stellar wind \citep{Whelan1973,Nomoto1982}. In the double degenerate (DD) scenario, two white dwarfs coalesce and ignite \citep{Webbink1984,Iben1984}. In both cases the strength of the radio emission depends on the density of the CSM which is in turn dependent on the density of the progenitor stellar wind. The shocked CSM can also produce X--ray emission which can be used to probe the CSM. \citet{Immler2006} claimed to have detected X--ray emission from the shocked CSM of the SNIa 2005ke, however this was not confirmed by \citet{Hughes2007}.

Many projects have searched for radio emission from SNeIa, but no detection has been made \citep{Eck2002,Eck1995,Weiler1989}. The largest set of radio observations of SNeIa is collected by \citet{Panagia2006}. In the SD scenario the current lack of a radio detection has placed constraints on the companion mass loss rate, and restricted the mass of a post main--sequence companion to be relatively low. At even lower luminosity and mass loss rates, the solely--SD scenario becomes untenable, whilst a radio detection of a SNIa would dis-confirm the solely--DD scenario which predicts no such emission. Radio observations of SNeIa can thus be used to determine whether or not all SNeIa result from only the SD or DD scenario, and can probe the ratio of SD to DD progenitors, in the case of a mixed population.

More sensitive measurements can be obtained from deeper observations, or via a combination of currently available observations. Stacking is the technique of combining observations or images to form a composite, and is the two dimensional equivalent of forming a weighted average. Stacking can been used to extract the average properties of a population of sources that are too faint to be detected individually and has been used at $\gamma$--ray \citep{Aleksic2011}, X--ray \citep{Laird2010}, optical \citep{White2007}, infrared and radio \citep{Garn2009} wavelengths.

This letter is organized as follows: \S\ref{sec:stacking} outlines the process of stacking and introduces visibility stacking, \S\ref{sec:targetselection} describes the data selection and reduction process, \S\ref{sec:results}--\ref{sec:discussion} present and discuss the results of the visibility stacking, and conclusions are drawn in \S\ref{sec:conclusions}.

\section{Visibility Stacking}\label{sec:stacking}
Traditionally stacking involves forming calibrated images of a single source which are then weighted and combined to produce a more sensitive image. For $N$ observations, each with an integration time of $t$, and Gaussian noise with an rms $\sigma$, a weighting scheme of $w=1/\sigma^2$ will result in a combined image with rms noise of $\sigma/\sqrt{N}$, which is equivalent to a single observation of total length $T=N\cdot t$. Such a stacking scheme is commonly used to break long observations into multiple shorter ones, to mitigate against problems such as radio interference, cosmic rays, and time--dependent calibration schemes, depending on the wavelength of observation. Non--Gaussian or correlated noise signals will result in a stacked image with a noise that is above the ideal $\sigma/\sqrt{N}$, and different weighting schemes are required to account for such problems.

Imaging radio synthesis observations involves a conversion between the visibility and image domains. The visibility sampling function (or $(u,v)$ coverage) of a radio interferometer is incomplete and generally non--uniform. Incomplete $(u,v)$ coverage will cause the reconstructed image to deviate from the true radio sky, with some spatial scales being under represented or absent. This problem is well known and motivates the design and layout of radio observatories \citep{Thompson1999} including the Very Large Array \citep[VLA,][]{Thompson1980}, Australia Telescope Compact Array \citep[ATCA,][]{frater_australia_1992}, the Low Frequency Array \citep[LOFAR,][]{Rottgering2003} and the Australian Square Kilometer Array Pathfinder \citep[ASKAP,][]{Johnston2009}. A common method for combating incomplete $(u,v)$ coverage is to combine observations that have complimentary $(u,v)$ coverages. For fixed arrays such as LOFAR and ASKAP this can be achieved with earth rotation synthesis, while movable arrays such as the VLA and the ATCA can also use multiple configurations. This commonly used process is the trivial case of stacking observations of a single object that are temporally separated.

Stacking of radio images of a population of sources can be done in the same manner as stacking at other wavelengths \citep[in the image domain,][]{Garn2009,Hodge2008,White2007}, however incomplete $(u,v)$ coverage can cause problems. Observations that are obtained from different configurations of a telescope, will have a different $(u,v)$ coverage and thus be sensitive to a different range of spatial scales and have different resolutions. An isolated point source can be easily detected at all spatial scales so stacking in the image domain will still recover the source. However, the noise statistics of the image is not easily understood. In more complicated fields, the distribution of sources and their morphologies become confused with spatial sampling effects and confident detections are difficult.

It is possible to benefit from improved $(u,v)$ coverage as well as the lower noise gained from a longer integration time by using a technique we term visibility stacking. Visibility stacking is achieved by combining visibilities from observations of a population of sources {\em before} the image is created.

When stacking in the image domain it is important to ensure that the input images are free of confusing sources, and the target source positions are aligned. The same is true for visibility stacking, but these corrections need to be applied to the visibility data. Removing confusing sources can be achieved by subtracting appropriate models from the visibility data. Aligning the target source positions can be achieved by adjusting the phases of the visibilities.

\section{Target selection}\label{sec:targetselection}
Observations of 27 nearby ($z<0.1$) SNeIa have been carried out by \citet{Panagia2006} using the VLA at wavelengths between 0.7 and 20~cm from 1981 to 2003. This data set has a large number of SNIa observations from a single instrument and a consistent set of calibrators making it ideal for forming a stacked image.

The data set is not completely homogeneous, with the VLA configuration being restricted by the observing season. The archival observations cover all standard configurations of the VLA (A, B, C, D, AnB, BnC, CnD), and are spread across multiple frequencies. In order to obtain the most sensitive stacked image, the 6\,cm observations were used as they constituted 75 of the 148 observations.

\begin{table*}
\caption{Archival VLA observations used in the stacking analysis.}
\label{tab:obs}
\begin{tabular}{clrrl|clrrl}
\hline
\hline
Name & Obs Date   & Age  & $\sigma_{6cm}$\uJy & dist &Name & Obs Date   & Age  & $\sigma_{6cm}$\uJy & dist \\
 SN  & Y-M-D      & days &  beam$^{-1}$       & Mpc  & SN  & Y-M-D      & days &  beam$^{-1}$       & Mpc  \\
\hline
1981B  & 1981-Mar-11 & 18   &  82 & 16.6   &1986O  & 1987-Feb-12 & 72   &  57 & 28     \\
\ditto & 1981-Apr-9  & 46   & 105 & \ditto &\ditto & 1987-Apr-11 & 130  &  68 & \ditto \\
\ditto & 1982-Jun-25 & 489  &  31 & \ditto &\ditto & 1987-May-24 & 173  &  67 & \ditto \\
\ditto & 1982-Oct-2  & 587  &  50 & \ditto &\ditto & 1987-Aug-28 & 269  &  52 & \ditto \\
1982E  & 1985-Dec-29 & 1417 &  44 & 23.1   &\ditto & 1989-Jul-17 & 958  &  90 & \ditto \\
1983G  & 1983-May-27 & 72   & 131 & 17.8   &1987D  & 1987-May-15 & 46   & 108 & 30     \\
1984A  & 1984-Mar-5  & 430  & 109 & 17.4   &\ditto & 1987-Jun-4  & 66   & 124 & \ditto \\
1985A  & 1985-Feb-1  & 49   &  93 & 26.8   &\ditto & 1987-Jun-21 & 83   &  32 & \ditto \\
\ditto & 1985-Feb-17 & 65   &  65 & \ditto &\ditto & 1987-Sep-18 & 172  &  62 & \ditto \\
\ditto & 1986-Jun-15 & 550  &  71 & \ditto &\ditto & 1988-May-29 & 426  & 174 & \ditto \\
\ditto & 1987-Oct-23 & 1045 & 104 & \ditto &1987N  & 1987-Dec-20 & 37   &  91 & 37.0   \\
1985B  & 1985-Feb-22 & 70   & 173 & 28.0   &\ditto & 1988-Feb-1  & 76   &  84 & \ditto \\
\ditto & 1985-Mar-18 & 94   &  50 & \ditto &\ditto & 1988-Mar-31 & 135  &  55 & \ditto \\
\ditto & 1985-Sep-15 & 275  &  74 & \ditto &\ditto & 1988-Aug-22 & 279  &  64 & \ditto \\
\ditto & 1986-Apr-30 & 502  &  81 & \ditto &1989B  & 1989-Feb-2  & 10   &  38 & 11.1   \\
\ditto & 1987-Sep-18 & 1008 &  56 & \ditto &\ditto & 1989-Feb-3  & 11   &  27 & \ditto \\
1986A  & 1986-Feb-7  & 18   & 126 & 46.1   &\ditto & 1989-Mar-6  & 42   &  39 & \ditto \\
\ditto & 1986-Feb-25 & 36   &  89 & \ditto &\ditto & 1989-Mar-27 & 63   &  57 & \ditto \\
\ditto & 1986-Mar-16 & 55   &  38 & \ditto &\ditto & 1989-Apr-6  & 73   &  64 & \ditto \\
\ditto & 1986-Apr-3  & 73   &  51 & \ditto &1989M  & 1990-Feb-13 & 271  &  64 & 17.4  \\
\ditto & 1986-Jun-15 & 146  &  67 & \ditto &\ditto & 1990-May-29 & 376  &  65 & \ditto \\
\ditto & 1987-Apr-1  & 436  &  68 & \ditto &1992A  & 1993-Feb-5  & 400  &  53 & 24.0  \\
       &             &      &     &        &1994D  & 1994-May-4  & 57   &  35 & 14    \\
       &             &      &     &        &1998bu & 1999-Jan-7  & 245  &  67 & 11.8  \\
%1986O  & 1987-Feb-12 & 72   &  57 & 28     \\
%\ditto & 1987-Apr-11 & 130  &  68 & \ditto \\
%\ditto & 1987-May-24 & 173  &  67 & \ditto \\
%\ditto & 1987-Aug-28 & 269  &  52 & \ditto \\
%\ditto & 1989-Jul-17 & 958  &  90 & \ditto \\
%1987D  & 1987-May-15 & 46   & 108 & 30     \\
%\ditto & 1987-Jun-4  & 66   & 124 & \ditto \\
%\ditto & 1987-Jun-21 & 83   &  32 & \ditto \\
%\ditto & 1987-Sep-18 & 172  &  62 & \ditto \\
%\ditto & 1988-May-29 & 426  & 174 & \ditto \\
%1987N  & 1987-Dec-20 & 37   &  91 & 37.0   \\
%\ditto & 1988-Feb-1  & 76   &  84 & \ditto \\
%\ditto & 1988-Mar-31 & 135  &  55 & \ditto \\
%\ditto & 1988-Aug-22 & 279  &  64 & \ditto \\
%1989B  & 1989-Feb-2  & 10   &  38 & 11.1   \\
%\ditto & 1989-Feb-3  & 11   &  27 & \ditto \\
%\ditto & 1989-Mar-6  & 42   &  39 & \ditto \\
%\ditto & 1989-Mar-27 & 63   &  57 & \ditto \\
%\ditto & 1989-Apr-6  & 73   &  64 & \ditto \\
%1989M  & 1990-Feb-13 & 271  &  64 & 17.4   \\
%\ditto & 1990-May-29 & 376  &  65 & \ditto \\
%1992A  & 1993-Feb-5  & 400  &  53 & 24.0   \\
%1994D  & 1994-May-4  & 57   &  35 & 14     \\
%1998bu & 1999-Jan-7  & 245  &  67 & 11.8   \\
\hline
\end{tabular}
\end{table*}

\subsection{Data reduction}
The raw data were obtained from the VLA archive\footnote{https://archive.nrao.edu/archive/advquery.jsp}, converted into uvfits format with {\sc AIPS} \citep{Greisen2003}, and reduced in {\sc MIRIAD} \citep{sault_retrospective_1995}. Flagging was based on manual inspection. In keeping with the analysis of \citet{Panagia2006}, 3C286 was used as a primary calibrator for each of the observations. The flux of the primary calibrator varied by less than 1\% over the course of the observing period.

A number of observations that were listed by \cite{Panagia2006} were not used in the final stacked image due to problems with: data access (4 could not be located in the VLA archive), data quality (6 had poor or missing calibrator observations, 4 had incorrect gains), or complicated field sources (15 observations, 8 of which include Centaurus A). The total number of usable observations was thus reduced to 46, and are listed in Table~\ref{tab:obs}.

In many cases the field of view of the VLA included background radio sources that created side--lobes and artefacts at the expected SNIa position which needed to be removed. A model of the sources was removed from affected visibilities. Column 4 of Table~\ref{tab:obs} lists the rms noise of individual observations when naturally weighted and with background sources removed.

Visibility stacking was achieved in {\sc MIRIAD} by adjusting the phase center of each image to be at the coordinates of the target SN, and imaging with the task {\sc invert}. Many of the observations contained two 50~MHz bands, 4.885 and 4.835\,GHz (and one observation at 4.889\,GHz), each of which were calibrated separately. A total of 79 such bands were used in the creation of the stacked images.

\begin{table*}[bht]
\centering
\caption{Parameters for the stacked images. Radio luminosity and inferred mass loss rates are $3\sigma$ upper limits.}
\label{tab:results}
\begin{tabular}{lcccc}
\hline
\hline
                       & \multicolumn{3}{c}{Image Stack} \\
                       & Early        & Late         & All          \\
\hline				      		                   
$d_o$ (Mpc)            & 14.3         & 19.0         & 15.7 \\
Effective Age (Days)   & 51.9         & 478          & 251          \\
Effective Integration  & \multirow{2}{*}{22.5} & \multirow{2}{*}{19.6} & \multirow{2}{*}{42.1} \\
at 50~MHz b/w (hr)        &\\ 
$1\sigma$ (\uJy beam$^{-1}$)  & 16.7         & 19.2         & 13.3         \\
Synthesized beam (arcsec$^2$) & $21\times18$ & $22\times15$ & $21\times17$\\
$L_o(\ergshz)$        & $1.2\e{25}$  & $2.5\e{25}$  & $1.2\e{25}$ \\
$\dot{M}(\moyr)$      & $1.3\e{-7}$  & $1.5\e{-6}$  & $5.4\e{-7}$  \\
\hline
\hline
\end{tabular}
\end{table*}

\subsection{Interpretation}
Images created via visibility stacking require some interpretation. In order to retain the greatest image sensitivity, the visibilities were naturally weighted when imaging. The flux in the image can thus be interpreted as being from a source at an effective distance $d_0$, which is a weighted average of the input source distances. The weighting function for $d_0$ is then the same as for the visibility data:

\begin{equation}
\frac{1}{d_0^4}=\frac{\frac{N_1}{d_1^4}+\frac{N_2}{d_2^4}+\dots+\frac{N_m}{d_m^4}}{N_1+N_2+\dots+N_m}
\label{eq:dzero}
\end{equation}

where $N_i$ is the number of visibilities in $i^{th}$ observation, and $d_i$ is the distance to the corresponding source. The function is of the form $N_i/d_i^4$ since doubling the integration time doubles $N_i$ and increases the sensitivity by $\sqrt{2}$ which allows a source of the same luminosity to be seen at a distance $\sqrt[4]{2}$ times further away.

Using Equation~(\ref{eq:dzero}), it is possible to calculate the effective distance probed by a stacked image, and convert the flux density of a detection or upper limit into an ensemble luminosity or luminosity limit. Distances listed in Table~\ref{tab:obs} are from \citet{Panagia2006}, and were used to calculate the effective distance $d_0$ for the stacked image.

A similar weighting function was applied to the age of a SN at the time of its observation to obtain an effective age that corresponds to the luminosity calculated in the above--mentioned manner. The relevant weighting function is:
\begin{equation}
\mathrm{Age}=\frac{\mathrm{Age}_1\cdot N_1+\mathrm{Age}_2\cdot N_2+\dots+\mathrm{Age}_m\cdot N_m}{N_1+N_2+\dots+N_m}
\label{eq:days}
\end{equation}

\section{Results}\label{sec:results}
\begin{figure*}[hbt]
\centering
\includegraphics[height=0.48\linewidth,angle=-90,bb=240 110 560 490]{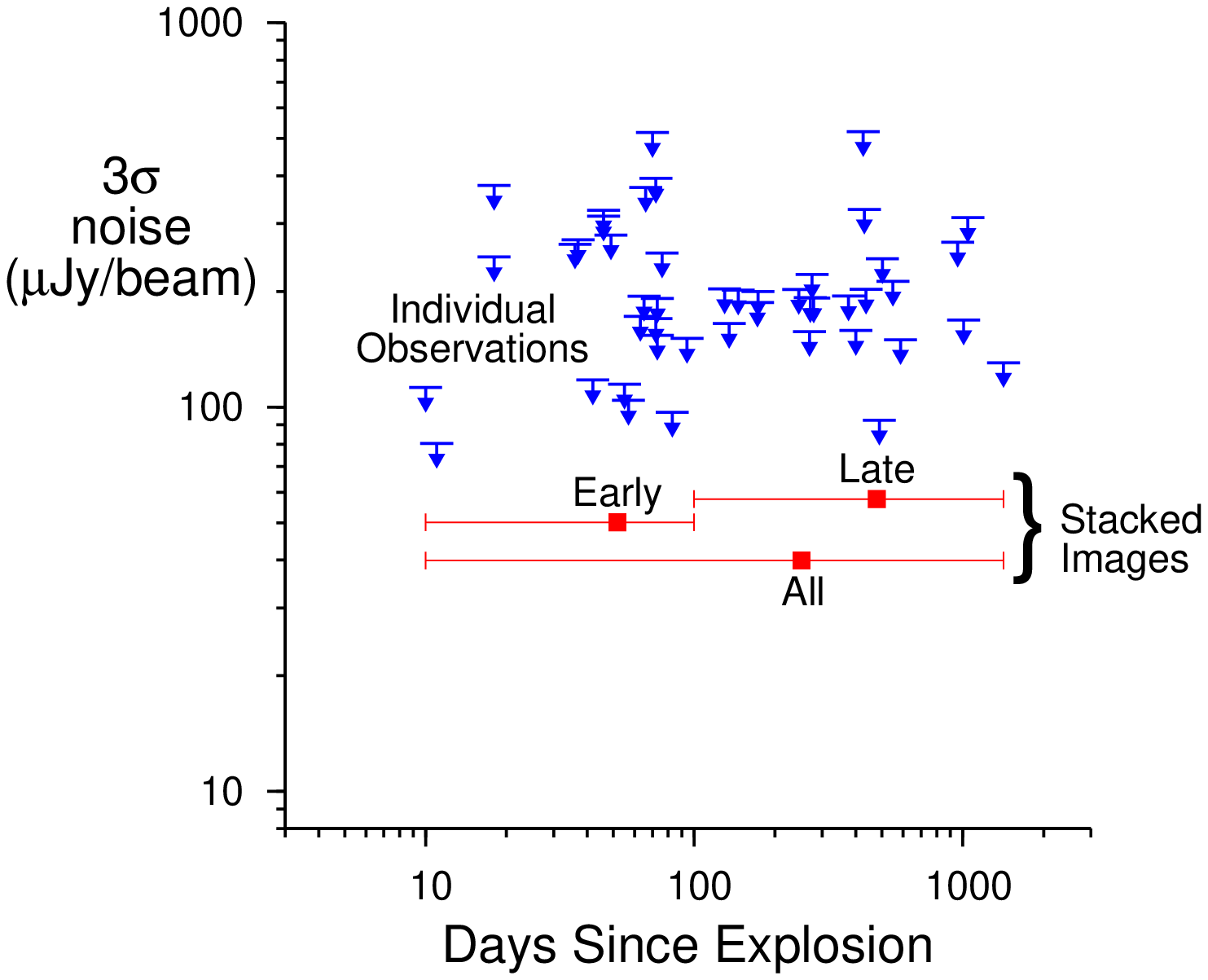}%figs/flux_time_plot.eps}
\includegraphics[height=0.48\linewidth,angle=-90,bb=240 110 560 490]{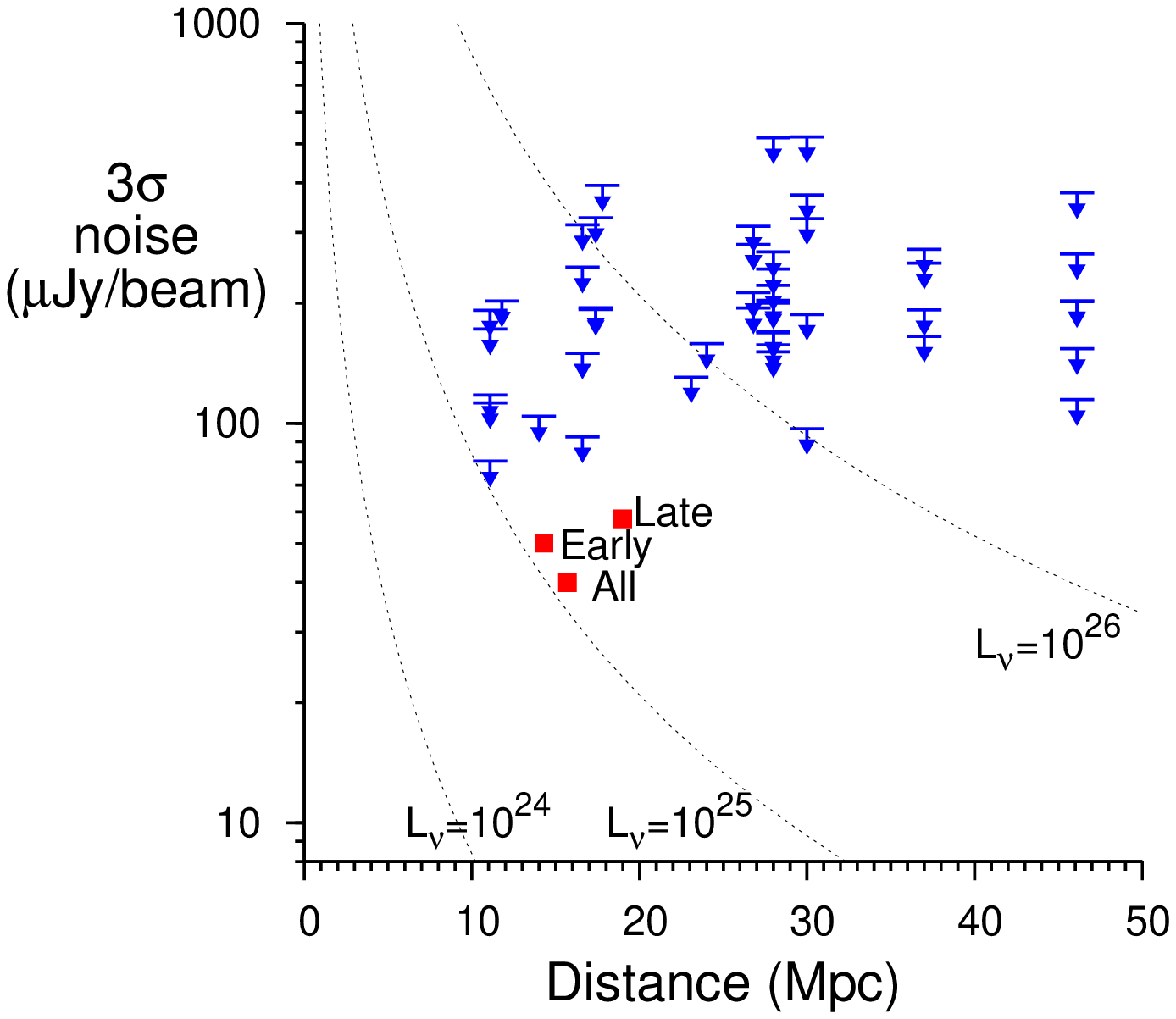}%figs/flux_distance_plot.eps}
\caption{The $3\sigma$ noise of the individual and stacked images as a function of SN age and distance. Three stacked images were created -- Early, All and Late. In each stacked image the noise level is fainter than in the individual observations. {\em Left:}  The noise in each image as a function of the age of the SNe at the time of observation. {\em Right:} The noise in each image as a function of SN distance. Lines have been drawn to show the $3\sigma$ upper limit for different 5\,GHz luminosities in \ergshz.}
\label{fig:flux_plots}
\end{figure*}

The 46 usable observations were combined to make three stacked images of the SNe. Two images were formed by selecting only observations with an age of either less than 100~days (Early) or greater than 100~days (Late), respectively. A third image was created using all of the available observations (All). In all three images no emission was seen above $3\sigma$ at the location of the SNe. The representative distance, age and corresponding luminosity of each of these images are listed in Table~\ref{tab:results}, and are plotted in Figure~\ref{fig:flux_plots} along with the upper limits obtained from the individual observations.

The stacked image of all observations represents a 50~MHz band width equivalent integration time of 42.1~hours, longer than any previous radio observation of any individual SNIa.

The final stacked images were not completely without structure and some faint sources were detected away from the expected SN location. Stacked images were created for each of the host galaxies, and faint sources that were not able to be seen or removed from the short observations were detectable in the more sensitive stacked image. These sources were found to be HII regions within, and the core of, NGC~3627 and were previously noted by \citet{Eck2002}.

Following \citet{Panagia2006, Panagia2011}, we model the luminosity of the SN1a as:

\begin{align}
\frac{L_\nu}{10^{26}\ergshz}=\Lambda\left(\frac{\dot{M}/10^{-6}\moyr}{w_{\mathrm{wind}}/10\kms}\right)^{1.65}\nonumber\\
\times\left(\frac{\nu}{5\mathrm{GHz}}\right)^{-1.1}\left(\frac{t}{1\,\mathrm{day}}\right)^{-1.5}e^{-\tau_{CSM}},
\label{eq:abslum}
\end{align}

where $t$ is the time since explosion, or age of the SN. The parameter $\Lambda$ is a dimensionless constant. On the assumption that the light curves of SNeIa are similar to that of SNeIb/c, we take $\Lambda=1285\pm 245$ as calculated by \citet{Panagia2006}. The value of $\Lambda$ can vary by a factor of 10 depending on which set of SNe are used in the derivation, however the dependence of $L_\nu$ on $\mdot/w_{\mathrm{wind}}$ results in an uncertainty in $\mdot$ of less than a factor of 2. A wind velocity $w_{\mathrm{wind}}=10\kms$ was assumed, consistent with a post main--sequence companion in the SD scenario. This model assumes that the CSM has a radial density profile $\rho\propto r^{-2}$.

The optical depth $\tau_\mathrm{CSM}$ of the CSM external to the blast wave depends on absorption mechanism (free--free or synchrotron self absorption), as well as the structure of the CSM (clumpy or uniform). Constant among all considerations of optical depth is that the luminosity peaks when the optical depth is unity. With the assumption $\tau_\mathrm{CSM}=1$, equation~(\ref{eq:abslum}) was used to estimate an upper limit for the mass loss rate $\dot{M}$, as listed in Table~\ref{tab:results}.

\section{Discussion}
\label{sec:discussion}
Combining data from different observations is a common technique for making measurements for a population of sources, however when dealing with upper limits care needs to be taken. A combined upper limit on the radio luminosity of SNeIa studied here can be obtained by creating a weighted average of the individual upper limits without stacking. Whilst this will result in a combined upper limit that is consistent with that obtained from a stacked image, it does not allow for the detection of sources. Stacking images permits the detection of source which would otherwise be below the detection limit, whilst combining upper limits implicitly assumes that no such detection will be made.

The stacked 5\,GHz images provide upper limits to the ensemble radio flux of SNeIa which are fainter than the individual observations. The stacked images of Early and All observations both have a $3\sigma$ upper limit on the 5\,GHz luminosity of $1.2\e{25}\ergshz$ which is the same as that of the best observation of SN1989B (1989-Feb-03). Whilst the stacked images are more sensitive than the individual observations (see Fig~\ref{fig:flux_plots} {\em left}), the fact that $L_\nu\propto\sigma\,d^2$ means that the flux/distance combination of the single observation and stacked image both correspond to the same luminosity limit ($1.2\e{25}\ergshz$ Fig~\ref{fig:flux_plots} {\rm right}). Rearranging Equation~\ref{eq:abslum} with $L_\nu\propto\sigma\,d^2$, we find that $\dot{M}\propto\sigma^{0.61}d^{1.2}t^{0.93}$, so that the limit on mass loss rate is more strongly dependent on the distance of the SN and age of observation than its sensitivity. Since the observation of SN1989B was made when it was only 10 days old, and the source is at only 11.1~Mpc, it provides a lower limit on the companion mass loss rate ($3.1\e{-8}\moyr$) than the older and more distant Early stacked image ($1.3\e{-7}\moyr$) even though the Early stacked image is more sensitive.

However, while the SN1989B observation provides an upper limit on SN luminosity which is the same as Early and All stacked images, it is only applicable to SN1989B and is not representative of the underlying population. If SNeIa are a completely homogeneous population of sources then this limit may be applied to all SNeIa, however the optical luminosity differences seen by \citet{Hamuy2000} and \citet{Timmes2003} suggest that this is not the case. By creating a stacked image of the SNeIa we are able to claim a luminosity limit which is applicable to all SNe within the sample, and be more confident that it is applicable to SNeIa as a whole. The $3\sigma$ luminosity limits and mass loss rates reported by \citet{Panagia2006}\footnote{Their $2\sigma$ upper limits have been converted to $3\sigma$ limits for comparison.} have a median of $1.2\e{26}\ergshz$ and $6.2\e{-7}\moyr$ respectively. The ensemble luminosity limits for the Early, Late and All stacked images are $5-10$ times lower these earlier limits, and the ensemble mass loss rate constraints of the Early and All stacked images are up to 5 times lower.

If the solely--SD scenario is correct then the mass transfer from the companion to the WD must result in some residual mass that is not accreted, our observations limit this contribution to the CSM to a rate of less than $1.3\e{-7}\moyr$. Wind accretion onto a binary companion is typically $\sim\,10\%$ efficient \citep{Yungelson1995}, implying a companion mass loss rate of $\leq 1.4\e{-7}\moyr$ ruling out red giants and medium to high mass ($2-10\msun$) post main--sequence stars as the possible binary companion. A main--sequence companion, having winds $\gg 10\kms$, could remain undetected even with much higher mass loss rates. Accretion via Roche--lobe overflow (RLOF) is more efficient than wind accretion and the models of \citet{Hachisu1999} predict that SNeIa formed from RLOF of a $1.3-2\msun$ red giant on to a CO white dwarf are consistent with the measured mass loss rate. The solely--SD scenario (with a red giant or post main--sequence companion $\geq 2\msun$) is thus not able to account for the properties of the observed SNIa population. A detailed analysis of the formation of binary systems with large mass ratios ($5-8\msun$ primary, $<2\msun$ secondary) is beyond the scope of this paper, but such systems may be relatively rare because of the amount of mass transfer from the primary to the companion during the final stages of the its evolution.

\section{Conclusions}\label{sec:conclusions}
By employing the new technique of visibility stacking we were able to create more sensitive images of SNeIa. The corresponding $3\sigma$ limits on the ensemble radio luminosity are considerably lower than most of the limits of \citet{Panagia2006}. Since the calculated mass loss rate depends on the distance and age of the SN as well as the sensitivity of the observation, we were unable to obtain an ensemble limit on $\dot{M}$ that was lower than the individual observation of SN1989B on Feb 3$^{rd}$ 1989. However, we were able to place an overall stringent limit on the ensemble 5\,GHz radio luminosity and mass loss rate for the SNIa population of $1.2\e{25}\ergshz$ and $1.3\e{-7}\moyr$ respectively.  By creating visibility stacked images of SNeIa we conclude that the solely single degenerate scenario (with a red giant or post main--sequence companion $\geq 2\msun$) cannot account for the observed SNIa population.

Visibility stacking is a powerful tool for analyzing radio observations of transient sources. The use of visibility stacking can be extended to the radio properties of any sample of radio quiet or radio faint sources, where image domain stacking has previously been used. The increased sensitivity can increase the confidence with which detections and upper limits are made. This technique needs to be applied carefully in the analysis of transient sources when such phenomena may result from more than one progenitor scenario.

In the era of time domain astronomy there will be a focus on obtaining more sensitive measurements on populations of sources that are transient. Such sources do not afford the luxury of repeated observations and thus techniques such as visibility stacking will be important in investigating source populations.

In studies which hope to detect the radio emission of SNeIa by increasing the sensitivity of the observations \citep[eg,][]{ChomiukLaura2011}, the ability to form visibility stacked images will further increase the sensitivity and applicability of the observations.

\section*{Acknowledgments}
The National Radio Astronomy Observatory is a facility of the National Science Foundation operated under cooperative agreement by Associated Universities, Inc. This research has been supported by the Australian Research Council through grants DP0987072 and FS100100033, and through the Science Leveraging Fund of the New South Wales Office for Science and Medical Research. The Centre for All--sky Astrophysics is an Australian Research Council Centre of Excellence, funded by grant CE11E0090.

{\it Facilities:} \facility{VLA}

\bibliographystyle{apj}
%\bibliography{ms} %use either this line or the following for the bibliography

\end{document}